\documentclass[%
 reprint,
 amsmath,amssymb,
 aps,
 prd,
floatfix,
twocolumn
]{revtex4-2}

\usepackage[usenames]{color}
\usepackage{amsmath}
\usepackage{graphicx}
\usepackage{dcolumn}
\usepackage{bm}

\begin{document}


\title{Effects of beta function on mass and melting temperature for scalar glueballs in AdS/QCD models at finite temperature}

\author{Alfredo Vega}
\email{alfredo.vega@uv.cl}
\affiliation{%
 Instituto de F\'isica y Astronom\'ia, \\
 Universidad de Valpara\'iso,\\
 A. Gran Breta\~na 1111, Valpara\'iso, Chile
}

\author{Amanda Rodriguez}%
 \email{amandarodriguezaburto@gmail.com}
\affiliation{%
 Instituto de F\'isica y Astronom\'ia, \\
 Universidad de Valpara\'iso,\\
 A. Gran Breta\~na 1111, Valpara\'iso, Chile
}

\date{\today}

\begin{abstract}

We consider an extension to finite temperature in an AdS/ QCD model, which regards anomalous dimension contributions to get a bulk mass depending on beta function. We study the effect of beta function on mass as a function of temperature and melting temperature for scalar glueballs.

\end{abstract}

\maketitle


\section{\label{intro}Introduction}

The emergence of AdS/CFT correspondence and its later developments have made it possible in recent decades to create models based on holographic correspondence in order to study different strongly coupled systems. In this sense, perhaps the most studied system considering holographic tools are hadrons, where different properties have been studied in a vacuum at zero temperature (e.g., see \cite{Vega:2010ns, Braga:2011wa, Brodsky:2006uqa}, or at finite temperature (e.g., see \cite{Colangelo:2009ra, Cui:2014oba, Fujita:2009wc, MartinContreras:2021bis, Gutsche:2019pls, Gutsche:2019blp}), and/or in dense media (e.g., see \cite{Stoffers:2010sp, Alho:2013hsa}).

The approaches considered use gauge / gravity ideas to study the properties of hadrons include the bottom-up approach, where starting from the properties of the hadrons, phenomenological models are built in an asymptotically gravity framework in
AdS spaces with additional dimensions. In this model, it is common to try to catch, in specific metrics, the environmental properties where hadrons are located (vacuum at zero temperature considering asymptotically AdS metrics and finite temperature and/or dense media with an AdS black hole metric), and introducing a hard cutoff in a holographic coordinate (hard wall models \cite{Erlich:2005qh, deTeramond:2005su}) or a soft cutoff with a non-dynamical dilaton (soft wall model \cite{Karch:2006pv}) to obtain a discrete mass spectrum for hadrons at zero temperature.

In the beginning, soft wall models were considered at zero temperature with an AdS metric and quadratic dilaton (e.g., see \cite{Karch:2006pv, Colangelo:2007pt, Brodsky:2007hb}). This was an advance in relation to hard wall models because in this case the model produced linear Regge trajectories for the mass spectrum of light mesons, but later it was necessary to introduce improvements to describe other hadron properties, or to include baryons correctly. Common extensions to soft wall models considers different asymptotically AdS metrics and/or dilatons beyond the quadratic one (e.g., see \cite{Forkel:2007cm, FolcoCapossoli:2019imm, Braga:2019yeh, Li:2013oda, MartinContreras:2020cyg}), but some authors have also considered bulk masses depending on the holographic coordinate (e.g., see \cite{Vega:2008te, Vega:2010ne, Vega:2011tg, Forkel:2010gu, Abidin:2009hr, Gutsche:2012bp}). The later alternative has been particularly useful in soft wall models to study barionic sector \cite{Abidin:2009hr, Gutsche:2012bp}, because in this case the dilaton is decoupled in equation of motions \cite{Kirsch:2006he}, rendering the introduction of any dilaton to reproduce the spectrum futile. 

A reason to modify masses for AdS modes propagating in bulk can be bound by using the AdS/CFT dictionary used in such models, where the bulk mass and operator dimension which create hadrons are related. Considering a classical dimension, the bulk mass is a constant characteristic of each hadron, but in terms of anomalous dimensions, this mass is a function of the holographic coordinate, due to anomalous dimensions depending on scale of energy, and this, according to the dictionary is related to the inverse of the holographic coordinate. The notion of considering an anomalous dimension to introduce a dependence on the holographic coordinate in bulk mass in some sense could be an excuse to introduce an extra term into the equation of motion, because with the exception of the authors of \cite{Boschi-Filho:2012ijd, FolcoCapossoli:2016uns}, no one can calculate this mass directly from anomalous dimensions.  

In \cite{Boschi-Filho:2012ijd}, the authors consider the full dimension for scalar glueballs, which is obtained from trace anomaly of energy-momentum tensor in QCD, which produces a dimension corresponding to a sum of the classical dimension plus an anomalous dimension, which depends on beta function. In their model, which considers an AdS metric and a quadratic dilaton, they improve the description of the scalar glueball mass spectrum at zero temperature. Throughout this paper, we consider the extension to the finite temperature region, and we explore effect of beta function on the behavior of masses as a function of temperature and melting temperature for scalar glueballs.

Before ending this introduction, it is important to mention that prior to \cite{Boschi-Filho:2012ijd}, there was a different approach, useful for holografic models different, which considered the use of the beta function to build holographic models \cite{Gursoy:2007cb, Gursoy:2007er, Gursoy:2010fj}.

The work consists of the following parts. Section II is a brief description of the soft wall model at zero temperature used to study scalar glueballs, where first we consider the most common approach, which ignores effects of beta function. Then, we discuss changes in bulk mass induced by anomalous dimensions, and their relation to beta function. In Section III, we discuss the extension to finite temperature, focusing on changes in scalar glueball masses as a function of temperature and melting temperature for these hadrons in three models, with and without beta function. Section IV provides conclusions and final comments for this work.

\section{Model at zero temperature}

\subsection{Model without beta function}

We consider a scalar field in a curved 5D space with a dilaton, whose action is given by

\begin{equation}
S=\frac{1}{2\,K}\int d^{5}x \,\sqrt{-g}\,e^{-\phi \left( z\right) } \mathcal{L},
\end{equation}
where 

\begin{equation}
\mathcal{L} = 
g^{MN}\,\partial _{M} \psi \left(x,z\right)\, \partial _{N} \psi \left( x,z\right)
+m_{5}^{2} \psi^{2}\left( x,z\right),
\end{equation}
with $M,N=0,1,2,3,z$, $z$ correspond to the holographic coordinate and $\phi\left(z\right)$ is a dilaton field. On the other hand, $m_{5}$ is the mass for modes propagating along the bulk, which according to the AdS/CFT dictionary in asymptotically AdS spaces is related to the operator dimension that create hadrons. This relation for scalars is \cite{Aharony:1999ti}:

\begin{equation}
    m_{5}^{2}R^{2}=\Delta \left(\Delta - 4\right),
\end{equation}
where $R$ is the $AdS$ radius (without  loss  of  generality we can consider this radius equal to one). For scalar glueballs $\Delta=4$ \cite{Boschi-Filho:2005xct, Colangelo:2007pt, Vega:2008af}.

The general metric considered in these models is 

\begin{equation}
    d^{2}s=e^{2 A(z)}\left(z\right)\eta_{MN}dx^{M}dx^{N},
\end{equation}
where $e^{2 A(z)}$ is a warp factor and $\eta_{MN}=diag(-1,1,1,1,1)$.

From this action we obtain an equation of motions in 5 dimensions for scalars, and we use the transformation

\begin{equation}
    \psi\left(X,z\right)=e^{-iPX}f\left(z\right),
\end{equation}
where $P$ and $X$ correspond to momentum and position in a 4D boundary. With the last transform we obtain an equation for $f(z)$, which depends on the holographic coordinate,

\begin{equation}
    -f''\left(z\right)+ B'(z) f'\left(z\right)+e^{2 A(z)}m_{5}^{2}f\left(z\right)=M^{2}f\left(z\right),
\end{equation}
where $B(z) = \phi(z) - 3A(z)$ and  $P^{2}=M^{2}$, i.e., $M$ is the mass of the hadron studied in these kinds of models.

Considering the transformation

\begin{equation}
    f\left(z\right) = e^{\frac{1}{2}B(z)} \psi\left(z\right),
\end{equation}
we obtain a Schr\"odinger equation like $\left(-\partial_{z}^{2}+V\left(z\right)\right)\psi\left(z\right)=M^{2}\psi\left(z\right)$, the potential of which, in terms of $A(z)$, $\phi(z)$ and $m_{5}$, is

\begin{equation}
\begin{split}
V(z) & = m_{5}^{2} e^{2 A(z)} + \frac{9}{4} (A'(z))^{2} - \frac{3}{2} A'(z) \phi'(z) \\
& ~~~~ + \frac{1}{4} \phi'^{2}(z) + \frac{3}{2} A''(z) - \frac{1}{2} \phi''(z).
\end{split}
\end{equation}

Considering an AdS metric  ($A(z) = Ln (1/z)$) and a quadratic dilaton $\phi(z) = c z^{2}$, the potential in the Schr\"odinger-like equation is

\begin{equation}
\label{Potencial}
    V\left(z\right)=\frac{15}{4z^{2}}+\frac{m_5^{2}}{z^{2}}+c^{2}z^{2}+2c.
\end{equation}

This choice of metric and dilaton throughout this work is called ``Model 1" \cite{Karch:2006pv, Colangelo:2007pt}. It corresponds to the first holographic soft wall model, which reproduces linear Regge trajectories in the light sector for mesons. It was very popular to study light hadrons, but for several years other metrics and dilatons have been studied to improve the phenomenology associated whith these kinds of models (e.g., see \cite{Braga:2019yeh, Li:2013oda, MartinContreras:2020cyg, Vega:2016gip, Gherghetta:2009ac}).

In Model 1 the only one parameter is $c = 0.2325~GeV^{2}$, which is fixed from $\rho$ masses \cite{Karch:2006pv}.

\begin{table}[t]
    \begin{tabular}{||c||c|c|c|c||}
    \hline
    \hline
    \textbf{$n$} & \textbf{Lattice \cite{Meyer:2004gx}} & \textbf{Model 1} & \textbf{Model 2} &  \textbf{Model 3}  \\
    \hline
    \hline
       1 & $1.475$ & $1.364$ & $1.497$  & $1.484$ \\
       2 & $2.755$ & $1.670$ & $2.307$  & $2.434$ \\
       3 & $3.370$ & $1.929$ & $3.056$  & $3.346$ \\
       4 & $3.990$ & $2.156$ & $3.625$  & $4.239$ \\
       \hline
       \hline
    \end{tabular}
    \caption{Scalar glueball ($J^{PC}=0^{++}$) mass spectra in GeV.}
    \label{tab:one}
\end{table}    

\subsection{Model with beta function}

As mentioned in the introduction, the AdS/CFT dictionary establishes a connection between bulk mass $m_{5}$ and the operator dimension that creates hadrons, and through this, it is possible to specify the hadron class we wish to study with our AdS/QCD model. The usual formula considers that $\Delta$ is the classical dimension for operators, but it is well known by which quantum effects these operators can develop anomalous dimensions that change with the energy scale. In this respect, in holographic correspondence and taking into consideration that the holographic coordinate is related to the energy scale, some authors have suggested that anomalous dimensions can modify bulk mass to produce a $z$-dependent mass and introduce improvements in holographic soft wall models with this mechanism (e.g., see \cite{Cherman:2008eh, Vega:2008te, Vega:2011tg, Forkel:2010gu}).

Among AdS/QCD models which include an anomalous dimension, the model presented in \cite{Boschi-Filho:2012ijd} is especially interesting to us, where the authors consider the full dimension for a scalar glueball extracted from the trace anomaly of the QCD energy-momentum tensor \cite{Narison:1988ts}. This, they obtain

\begin{equation}
\Delta = 4 + \beta'(\alpha) - \frac{2}{\alpha} \beta(\alpha)
\end{equation}
which in terms of 't Hooft coupling $\lambda = N_{c} g_{YM}^{2} = 4 \pi N_{c} \alpha$ produces

\begin{equation}
\Delta = 4 + \beta'(\lambda) - \frac{2}{\lambda} \beta(\lambda),
\end{equation}
where the prime denotes the derivative respect $\lambda$ and 

\begin{equation}
\mu \frac{d \lambda(\mu)}{d~\mu} = \beta (\lambda(\mu)).
\end{equation}

\begin{figure}
  \includegraphics[width=3.1 in]{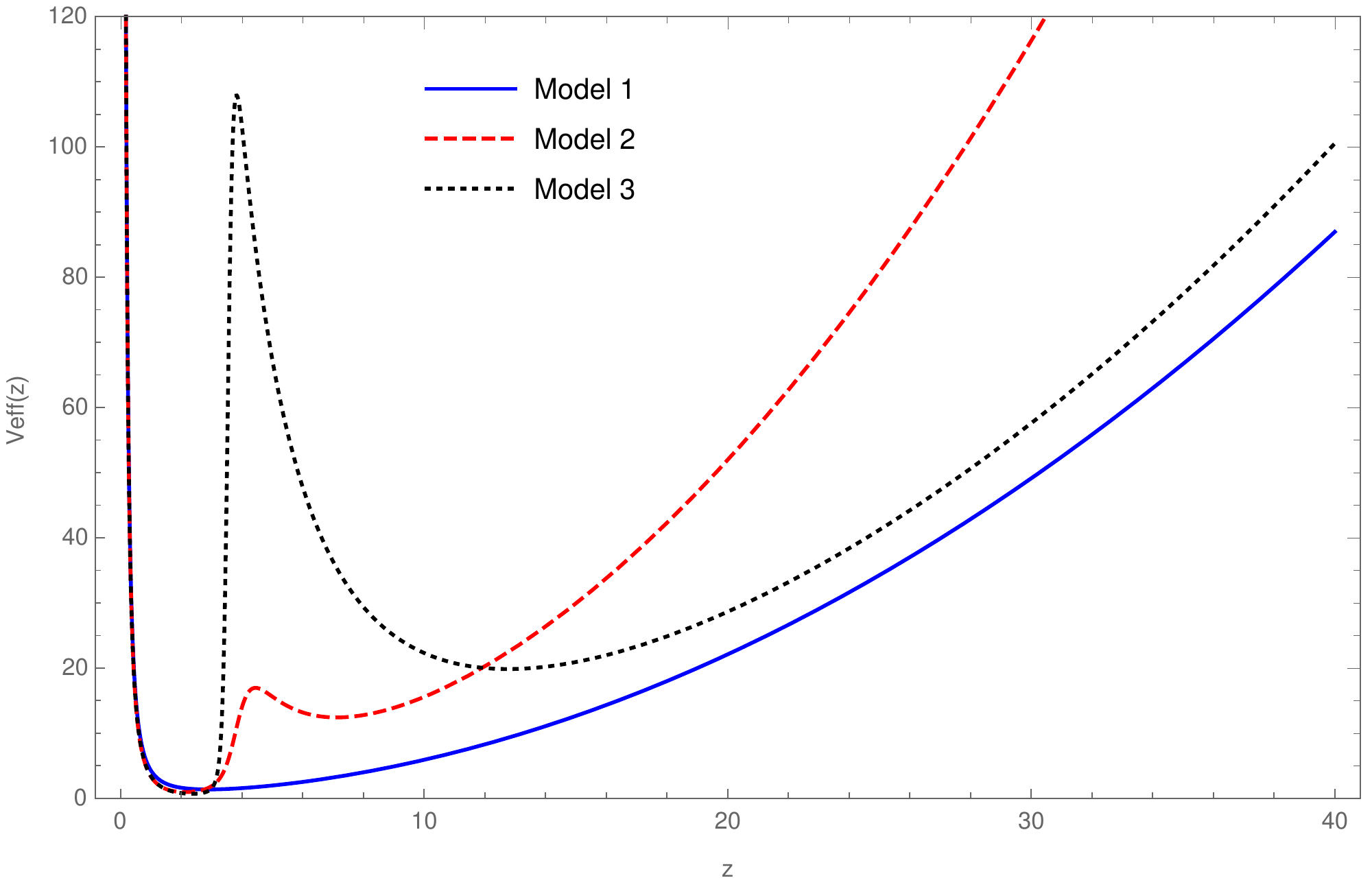}
\caption{Potential for scalar glueball in three models Potentials considered in this work.}
\label{fig:three}
\end{figure}

Remembering that the holographic coordinate is related to $\mu^{-1}$, where $\mu$ is the scale of the renormalization group, it is possible to write the equation for beta function in terms of $z$ and obtain

\begin{equation}
\label{EqBeta}
z \frac{d \lambda(z)}{d~z} = - \beta(\lambda(z)).
\end{equation}

This last expression shows that considering an anomalous dimension for scalar glueballs produces a $m_{5}$ that depend on $z$, which specifically depends on the beta function according to

\begin{equation} \label{eq1}
\begin{split}
m_{5}^{2}(z) & = \Delta(z) (\Delta(z) - 4) \\
 & = \biggr( 4 + \beta'(\lambda) - \frac{2}{\lambda} \beta(\lambda) \biggl) \biggr( \beta'(\lambda) - \frac{2}{\lambda} \beta(\lambda) \biggl).
\end{split}
\end{equation}

Replacing this result in (\ref{Potencial}), we obtain a potential which depends on the beta function.

Apart from Model 1, in this work we consider two other models studied in \cite{Boschi-Filho:2012ijd} in the zero temperature case. We consider the effect at finite temperature in ``Model 2"the beta function of which is

\begin{equation}
\beta(\lambda) = -b_{0} \lambda^{2} \biggr(1 - \frac{\lambda}{\lambda_{*}} \biggl).
\end{equation}

In this case, the solution for (\ref{EqBeta}) is

\begin{equation}
\lambda(z) = \frac{\lambda_{*}}{1 + W((z_{0}/z)^{b_{0} \lambda_{*}} (\frac{\lambda_{*} - \lambda_{0}}{\lambda_{0}}) e^{\frac{\lambda_{*} - \lambda_{0}}{\lambda_{0}}})},
\end{equation}
where $W(x)$ is the Lambert function and $\lambda(z_{0}) = \lambda_{0}$ determine the integration constant.

The parameters in this model are $b_{0}=11/24\pi^{2}$, $z_{0} = 1~GeV^{-1}$, $c = -0.36~GeV^{2}$, $\lambda_{*} = 350$ and $\lambda_{0} = 18.5$.

Additionally, we consider ``Model 3", with this beta function

\begin{equation}
\beta(\lambda) = - \frac{b_{0} \lambda^{2}}{1 + b_{1} \lambda}.
\end{equation}
Solving (\ref{EqBeta}), we obtain
\begin{equation}
\lambda(z) = \frac{1}{b_{1} W (\frac{e^{1/(b_{1} \lambda_{0})}}{b_{1} \lambda_{0}} (z_{0}/z)^{b_{0}/b_{1}}}.
\end{equation}

The parameters in Model 3 are $b_{0}=11/24\pi^{2}$, $z_{0} = 1~GeV^{-1}$, $c = -0.25~GeV^{2}$, $b_{1} = 1.3 * 10^{-3}$ and $\lambda_{0} = 19$.

Fig. 1 shows the potentials for all three models in this work, and Table 1 summarizes the masses of first scalar glueball states at zero temperature in each holographic model and includes a column with results from lattice \cite{Meyer:2004gx}.

\section{Model at finite temperature}

In this kind of model, thermal effects are captured mainly in metrics. Here, the metric considered is 
\begin{equation}
ds^{2}=e^{2\,A\left( z\right) }\left[ -f\left( z\right)
dt^{2}+\sum_{i=1}^{3}\left( dx^{i}\right) ^{2}+\frac{1}{f\left( z\right) }\,
dz^{2}\right],
\end{equation}
or
\begin{equation}
g_{MN}=e^{2\,A\left( z\right) }\text{ diag}\left( -f\left( z\right) ,\,1,\,1,\,1,%
\,\frac{1}{f\left( z\right) }\right).
\end{equation}

The equation of motion associated with this action is
\begin{widetext}

\begin{equation}
e^{B\left( z\right) }\,f\left( z\right) \,\partial _{z}\left[ e^{-B\left(
z\right) }\,f\left( z\right) \,\partial _{z}\psi \right] -f\left( z\right)\,
e^{2A\left( z\right) }\,m_{5}^{2}\,\psi +\omega ^{2}\,\psi -f\left( z\right)\,
q^{2}\,\psi =0,
\end{equation}
where $B(z) = \phi(z) - 3A(z)$, as in zero temperature case.

Considering  our particles at rest, we fix ($\vec{q} = \vec{0}$). Thus the previous equation looks like

\begin{equation}
\label{EOM}
\partial _{z}\left[ e^{-B\left( z\right) }\,f\left( z\right) \,\partial _{z}\psi %
\right] +\left[ \frac{\omega ^{2}}{e^{B\left( z\right) }\,f\left( z\right) }
-e^{-\phi \left( z\right) +5A\left( z\right) }\,m_{5}^{2}\right] \psi =0,
%
\end{equation}
\end{widetext}
where $\omega^{2}$ is related to the hadron mass in the thermal bath. Starting from the last equation, it is possible to obtain the spectral function, and with study the effect of the temperature on the hadron masses and find the melting temperature for the different species in this medium.

In terms of the variable $u$ defined by $z = z_{h} u$, the horizon is located in $u=1$, and the Green function can be calculated using solutions of (\ref{EOM}) such as, 
\begin{equation}
\label{Green-F-Ther}
G_R(\omega)=-\left.\frac{2}{z_h\,\mathcal{N}}\,\,e^{-B(u)}\,f(u)\,\psi(u,-\omega)\,\partial_u\,\psi(u,\omega)\right|_{u=1},
\end{equation}
which is related to the spectral function by 

\begin{equation}
\rho(\omega) = - Im~G_{R}(\omega).
\end{equation}

\begin{figure}[b]
  \includegraphics[width=3.1 in]{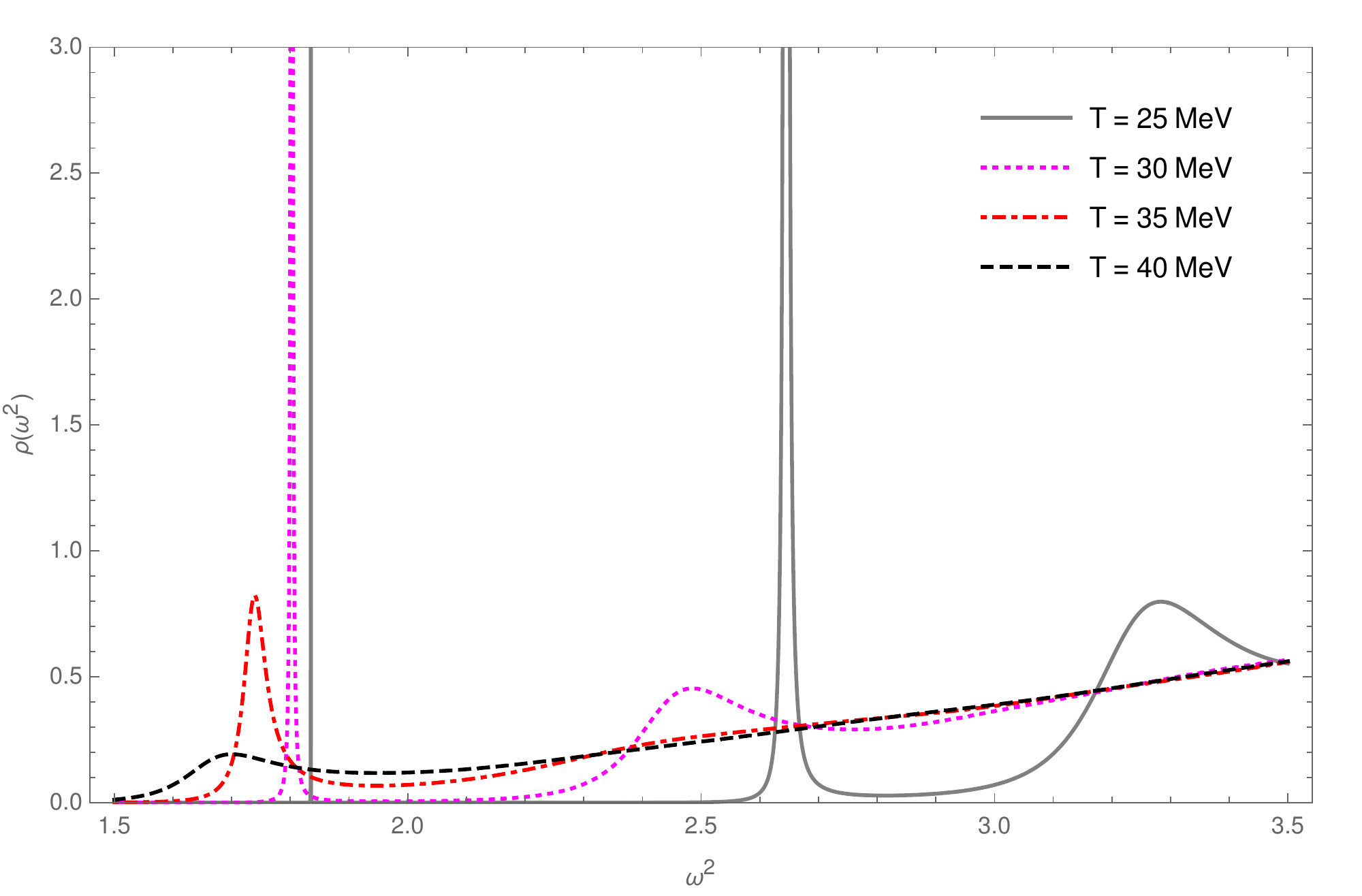}
\caption{Spectral function in Model 1 at different temperatures.}
\label{fig:three}
\end{figure}

Fig. 2 shows the spectral function in Model 1 calculated at different temperatures. We can see that in each case the spectral function presents several picks, the locations of which correspond to squared of hadron mass at this temperature. It is also possible to see that when the temperature increases, two things happen: on the one hand, the location of maximum change, and following these changes in temperature, we obtain variations of hadron mass with temperature. On the other hand, high of peaks it is reduced and these are wider up to disappear, which is interpreted as melting in thermal bath.

Although the spectral function carries information of state at any temperature, to calculate spectral function at low temperature could be numerically difficult because peaks are very high and narrow. However, fortunately, as was discussed in \cite{Colangelo:2009ra}, if it is possible to tackle the problem of obtaining masses by considering that the horizon is at a long distance from the border, and modes dual to hadrons decrease to zero far from the horizon, then it is also possible to solve (\ref{EOM}) considering normalized modes and using the usual tools to solve Sturm-Liouville problems. When horizon is close, and we cannot consider normalized modes, peaks are wide enough and we can perform an analysis considering Breit-Wigner functions to obtain information about masses at different temperatures and obtain the melting temperature.

The Breit-Wigner analysis mentioned in last paragraph considers that the spectral function has a structure like $\rho(\omega) = \rho_{BW}(\omega) + \rho_{B}(\omega)$, where $\rho_{BW}(\omega)$ is Breit-Wigner-like and $\rho_{B}(\omega)$ represents the background. As the first step, it is neccesary to subtract the background in the spectral function to isolate the resonance-like behavior in the spectral function. Alternatives to removing the background include different polynomials in $\omega^{2}$ (e.g., see \cite{Colangelo:2009ra, Cui:2014oba, Fujita:2009wc}), but here we follow the procedure suggested in \cite{MartinContreras:2021bis}, which consider a polynomial interpolation obtained with the Mathematica software. Later, with the resonance isolated, the usual methodology for making an adjustment with a Breit-Wigner function and extracting information about the location of maximum (to obtain the mass) and the width of the Breit-Wigner function, but here we follow a slightly different procedure, because numerically we calculate the location of maximum in $\rho_{BW}(\omega)$ and width directly, with the latter being defined as the distance between points, where $\rho_{BW}(\omega)$ is reduced to half. In addition to the location of maximum in resonance, which is related to the hadron mass at this temperature, with the height and width, a melting criteria is established for hadrons. There are several rules with regard to melting (e.g., see \cite{Colangelo:2009ra, Cui:2014oba, Fujita:2009wc}), and we use height divided by width equals one as the melting criterion, as in \cite{MartinContreras:2021bis}.

From an analysis of the spectral function according to the guidelines mentioned in the previous paragraph, in each model we calculate $m^{2}(T)$ for lighter scalar glueballs and their melting temperature. Figure 3 shows mass squared versus temperature in the three models included in this work, and we can see that in models which consider the effect of beta function, the melting temperature is increased and masses experience significant changes at high temperatures compared to Model 1, which disregards beta function effects.

Considering mass at zero temperature as the reference, in Model 1 mass is reduced by 8.4\% before melting, a reduction which is increased to 44.7\% in Model 2 and 63.6\% in Model 3, respectively. In relation to the melting temperature of our results, considering the subtraction procedure in the spectral funtion, the Breit-Wigner analysis and melting criteria are $T = 39\,MeV$ for Model 1, $T = 84\,MeV$ for Model 2, and $T = 95\,MeV$ for Model 3.

\begin{figure}
  \includegraphics[width=3.1 in]{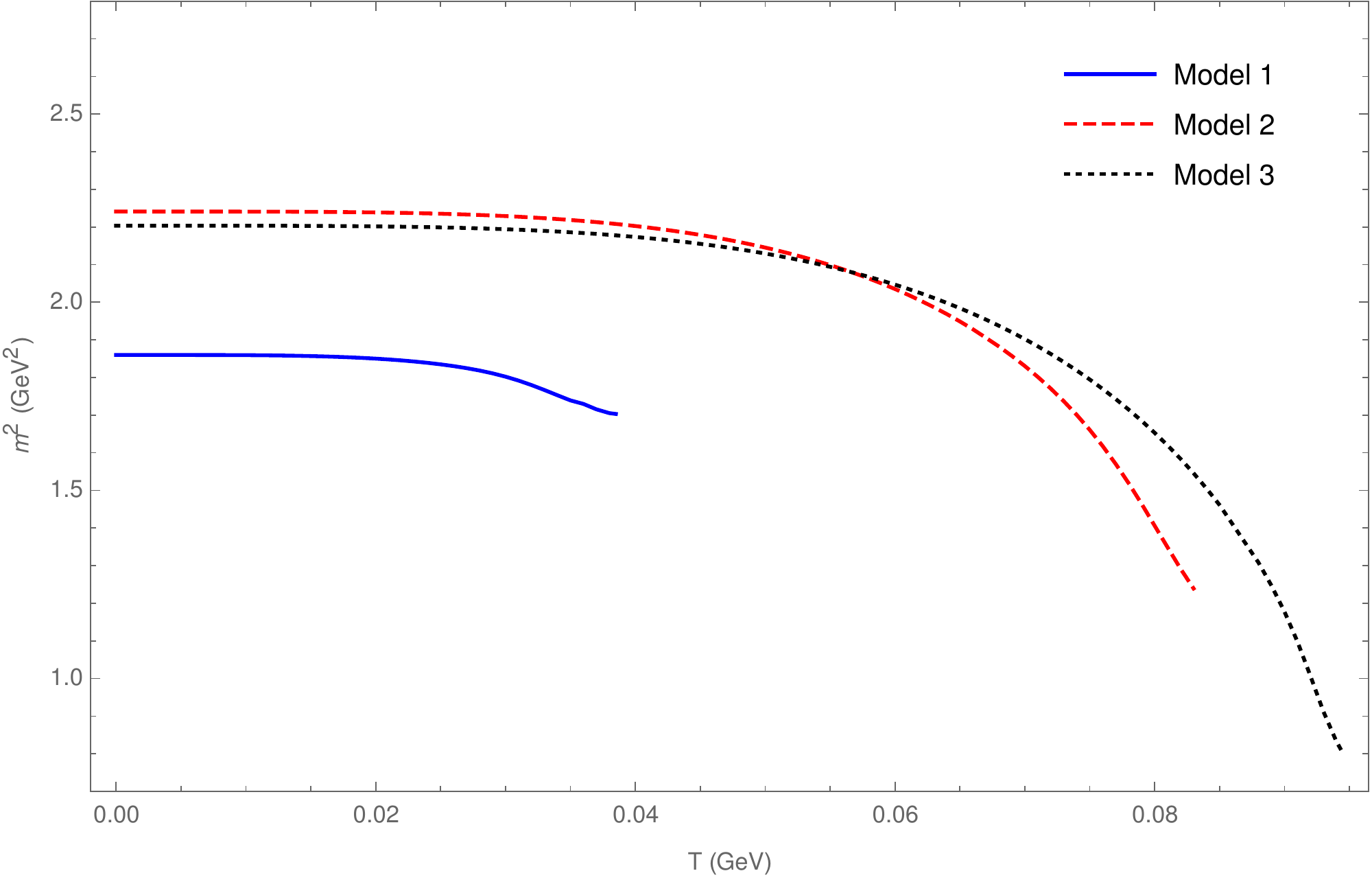}
\caption{$m^{2}~v/s~T$ for lighter scalar glueballs in each model studied in this paper.}
\label{fig:three}
\end{figure}

\section{Conclusions}

We performed an extension to finite temperature for an AdS/QCD model presented in \cite{Boschi-Filho:2012ijd}, which at zero temperature improved the spectra of scalar glueballs considering the effect of anomalous dimensions to modify bulk mass, making it dependent on beta function.

We notice that the incorporation of an anomalous dimension which depends on beta function in AdS/QCD models at finite temperature produces important changes in the hadron properties studied. First, the range of variation in the lighter scalar glueball mass is increased compared to the model with no beta function. As we mentioned in the last paragraph of the previous section, in Model 1 the ground state mass of the scalar glueball is reduced by 8.4\% before melting compared to its value at zero temperature, a reduction which is increased to 44.7\% in Model 2 and 63.6\% in Model 3, respectively. Additionally melting temperature in models with a beta function is increased more than double the value obtained in Model 1.

An increase in melting temperature was expected behavior according to the qualitative analysis based on studies of potentials, because the beta function produces a local maximum in potential in the zero temperature case. This maximum produces a potential at finite temperatures with a well able to contain quasibound states at higher temperatures, increasing the melting temperature, as was observed in our direct calculations based on spectral function.

\begin{acknowledgments}
We wish to acknowledge the financial support provided by FONDECYT (Chile) under Grants No. 1180753.
\end{acknowledgments}

\bibliography{references}

\end{document}